\documentclass[prd,twocolumn,nofootinbib,a4paper]{revtex4}

\usepackage{graphicx,epsfig,amssymb,subfigure}
\usepackage[latin1]{inputenc}

\def\be {\begin{equation}}
\def\ee {\end{equation}}
\def\ba {\begin{eqnarray}}
\def\ea {\end{eqnarray}}
\def\nn {\nonumber}
\def\bea{\begin{eqnarray}}
\def\eea{\end{eqnarray}}
\def\bi {\begin{itemize}}
\def\ei {\end{itemize}}

\begin{document}

\title{Area spectrum of rotating black holes via the new interpretation\\ of
quasinormal modes}

\author{Elias C. Vagenas}
\email{evagenas@academyofathens.gr}
\affiliation{Research Center for Astronomy \& Applied Mathematics\\
 Academy of Athens\\
 Soranou Efessiou 4\\
 GR-11527, Athens, GREECE}

\date{\today}

\begin{abstract}
Motivated by the recent work on a new physical interpretation of
quasinormal modes by Maggiore, we utilize this new proposal to the
interesting case of Kerr black hole. In particular, by modifying
Hod's idea, the resulting black hole horizon area is quantized and
the resulting area quantum is in full agreement with Bekenstein's result.
Furthermore, in an attempt to show that the area spectrum is
equally spaced, we follow Kunstatter's method. We propose a new
interpretation as a result of Maggiore's idea, for the frequency
that appears in the adiabatic invariant of a black hole. The
derived area spectrum is similar to that of the quantum-corrected Kerr black hole
but it is not equally spaced.
\\

\end{abstract}
\maketitle
\par\noindent
Since the onset of General Relativity black holes have been a
matter of major concern  for the scientific community. This
interest is twofold. On one hand, black holes are astrophysical
objects whose fingerprints will be observed on recent or future
detectors for gravitational waves e.g. LIGO
\cite{Abramovici:1992ah} and VIRGO \cite{Acernese:2002bw}. On the
other hand, black holes have always been a test bed for any
proposed scheme for a quantum theory of gravity. It is evident
that it would be of great importance for quantum gravity (and not
only) if the superficially distinct (astrophysical vs theoretical)
aspects could be reconciled. Hod was one of the first to make such
a phenomenological work \cite{Hod:1998vk}. He combined the
perturbations of astrophysical black holes with the principles of
Quantum Mechanics and Statistical Physics in order to derive the
quantum of the black hole area spectrum. Following this line of
thought, Kerr black holes are the most interesting black hole
solutions since from the astrophysical point of view are the most
important ones while from the purely theoretical point of view are
more complicated than the simple Schwarzschild black hole.The
metric of a four-dimensional Kerr black hole given in
Boyer-Lindquist coordinates is \bea
ds^{2}&=&-(1-\frac{2Mr}{\Sigma})dt^{2}-\frac{4Mar
\sin^{2}\theta}{\Sigma}dtd\varphi+\frac{\Sigma}{\Delta}dr^{2}\nn\\
&+&\Sigma d\theta^{2}+(r^2+a^2+2Ma^2r\sin^{2}\theta)
\sin^{2}\theta d\varphi^{2} \label{met} \eea where, as always, $M$
is the mass of the black hole, $J$ is the angular momentum of the
black hole,  $a$ is the specific angular momentum defined as
$J/M$, $\Sigma=r^2+a^2\cos^2 \theta$, and $\Delta=r^2-2Mr+a^2$.
The roots of $\Delta$ are given by \be r_{\pm}=M\pm
\sqrt{M^{2}-a^{2}} \label{root} \ee where $r_{+}$ is the radius of
the event (outer) black hole horizon and $r_{-}$ is the radius of
the inner black hole horizon.
The Kerr black hole is rotating with angular velocity (evaluated
on the event black hole horizon) \bea
\Omega&=&\frac{a}{r_{+}^{2}+a^{2}}=\frac{J}{2M\left(M^{2}+\sqrt{M^{4}-J^{2}}\right)}
\label{angve}~. \eea Furthermore, the horizon area and the Hawking
temperature of Kerr black hole (in gravitational units) are given,
respectively, by \bea A&=&4\pi
(r_{+}^{2}+a^{2})=8\pi\left(M^{2}+\sqrt{M^{4}-J^{2}}\right)
\label{area} \eea and \bea
T_{H}&=&\frac{r_{+}-r_{-}}{A}=\frac{\sqrt{M^{4}-J^{2}}}{4\pi
M\left(M^{2}+\sqrt{M^{4}-J^{2}}\right)} \hspace{1ex}.
\label{temeq} \eea

\par\noindent
As mentioned before, Hod managed to derive the quantum of the area
spectrum using the Bohr's Correspondence principle and the complex
spectrum of the quasinormal modes that correspond to the
perturbation equation of Schwarzschild black hole. The resulting
quantum was of the form \cite{Hod:1998vk} \be \Delta A = 4
l^{2}_{p} \ln 3 \label{hodquantum} \ee where $l_{p}$ is the Planck
length. Hod's idea\footnotemark[1]
\footnotetext[1]{Later, it was shown by Natario and Schiappa \cite{Natario:2004jd} that Hod's calculation
is not universal since it depends on the asymptotics of the black hole spacetime under study.} rejuvenated the interest of the research
community for the quantization of the black hole area spectrum and
subsequently for a derivation of black hole entropy from
Statistical Physics. Actually, the aroused interest was
strengthened by the possible links with loop quantum gravity as
proposed by Dreyer\footnotemark[2]
\footnotetext[2]{However, it should be emphasized that the method used by Dreyer for state
counting was incorrect and consequently, Dreyer computed an incorrect value for the Barbero-Immirzi parameter.
The correct method of counting states was proposed by Domagala and Lewandowski \cite{Domagala:2004jt}.
Furthermore, implementing the correct method Meissner calculated the correct value for  Barbero-Immirzi parameter
\cite{Meissner:2004ju} which was between the upper and lower bounds set in \cite{Domagala:2004jt}.}\cite{Dreyer:2002vy}.

\par\noindent
Some thirty five years ago, by proving that the black hole horizon
area is an adiabatic invariant, Bekenstein showed that the quantum
of black hole area is of the form \cite{Bekenstein:1974jk} \be
\Delta A =8 \pi l^{2}_{p}~. \ee Adiabatic invariants of a system
are quantities which vary very slowly compared to variations of
the external perturbations of the system. Moreover, given a system
with energy $E$ and vibrational frequency $\omega(E)$, one can
show that the quantity $E/\omega$ and therefore \be I=\int
\frac{dE}{\omega(E)}~, \label{adiabatic} \ee is an adiabatic
invariant. For the case of black holes, as already said above,
Bekenstein was the first to state that the adiabatic invariants
are the black hole horizon areas
\cite{Bekenstein:1997bt,Bekenstein:1998aw}.

\par\noindent
Exploiting the idea of adiabatic invariants and the statement by
Bekenstein \cite{Bekenstein:1974jk}, Kunstatter
\cite{Kunstatter:2002pj} derived for the $d(\geq 4)$-dimensional
Schwarzschild black hole an equally spaced entropy spectrum. Key
points to Kunstatter's approach were :
\\
(1) the first law of black hole thermodynamics which for the case
of a Schwarzschild black hole is of the form \be
dM=\frac{1}{4}T_{H}dA \label{sch1law}~, \ee
\\
(2) Hod's proposal that in the asymptotic limit, i.e. the large
$n$ limit, the real part of quasinormal frequencies of the
Schwarzschild black hole
uniquely fixes the quantum of the black hole area spectrum, and\\
(3) the fact that the Bohr-Sommerfeld quantization has an equally
spaced spectrum in the large $n$ limit, i.e. \be I \approx n\hbar
\label{smi}~. \ee Kunstatter viewed the Schwarzschild black hole
as a system whose adiabatic invariant takes the form
\be
I= \int
\frac{dM}{\omega_{R}}
\hspace{1ex}
\ee
where $dE$ was set equal to $dM$ and the frequency in the denominator of the integral in
equation (\ref{adiabatic}) was set equal to the real part of the
quasinormal frequencies of the Schwarzschild black hole which was
$\omega_{R} \sim T_{H}$. Finally, the area spectrum and thus the
entropy of the Schwarzschild black hole were discrete and equally
spaced. At that point Kunstatter  raised the interesting question
if the aforesaid derivation holds for  rotating black holes. In
this direction, Hod studied analytically the quasinormal modes of
Kerr black hole \cite{Hod:2003hn} and he concluded that the
asymptotic quasinormal frequencies of Kerr black hole are given by
the simple expression \be \omega=m\Omega - i 2\pi T_{H} n
\label{quasi1} \ee which were in agreement for the case
of $l=m=2$ with the numerical results derived by Berti and
Kokkotas \cite{Berti:2003zu}.\\

\par\noindent
Endeavoring to answer Kunstatter' question we extended his
approach  \cite{Setare:2004uu} to the case of Kerr black hole
using the real part of the quasinormal frequency given in equation
(\ref{quasi1}). The first law of black hole thermodynamics is now
written as \be dM=\frac{1}{4}T_{H}dA+\Omega dJ \label{kerr1law}
\ee where the angular velocity is given by equation (\ref{angve})
and obviously the corresponding expression for  adiabatic
invariant is now given by the expression
\be
I= \int
\frac{dM-\Omega dJ}{\omega_{R}}
\label{adiabkerr}
\hspace{1ex}.
\ee

\par\noindent
Equating Bohr-Sommerfeld quantization condition (\ref{smi}) with
the adiabatically invariant integral (\ref{adiabkerr}) one obtains
an area spectrum for the Kerr black hole which  although discrete,
is not equidistant.
However, it was proven by Bekenstein
\cite{Bekenstein:1974jk,Bekenstein:1997bt} and others
\cite{Makela:2000zd,Gour:2002ga} that the area spectrum of Kerr
black hole is  discrete and uniformly spaced. Therefore, it was
concluded that the function that was used in the above-mentioned
computation as real part of the asymptotic quasinormal frequencies
of Kerr black hole, i.e. expression (\ref{quasi1}), was not the
correct one. Recent analytical works \cite{Keshet:2007nv,Keshet:2007be}
confirmed older numerical calculations
\cite{Berti:2004um} in which  the quasinormal frequencies of a
Kerr black hole are of the form \be \omega(n)= \tilde{\omega}_{0}-
i \left[4\pi T_{0}\left(n+\frac{1}{2}\right)\right] \label{quasi2}
\ee where ${\tilde\omega_{0}}$ is a function of the black hole
parameters and $T_{0}$ is the effective temperature. For $M^{2}\gg
J$, or equivalently $a/M \approx 0$, the effective temperature is
\be T_{0}(a)\approx -\frac{T_{H}(a=0)}{2} \label{efftemp} \ee and
$T_{H}(a=0)$ is the Hawking temperature of the Schwarzschild black
hole (henceforth $T_{H}^{Sch}$). The subscripts of the frequency
$\tilde{\omega}$ and temperature $T$ in equation (\ref{quasi2})
denote that these quantities have been computed by integrating a
contour that crosses the real axis outside the event horizon
\cite{Keshet:2007be}.

\par
Very recently a new physical interpretation for the quasinormal
modes of black holes was given by Maggiore \cite{Maggiore:2007nq}.
According to Maggiore's proposal if one wants to avoid several
problems in the interpretation of quasinormal frequencies when
compared with macroscopical systems, one has to treat a perturbed
black hole as a damped harmonic oscillator. Then one has to
identify as proper frequency of the equivalent harmonic oscillator
the following quasinormal normal frequency
\be
\omega_{0}=\sqrt{\omega^{2}_{R}+\omega^{2}_{I}}
\label{quasi3}
\ee
which decidedly for the case of long-lived quasinormal modes, i.e.
$\omega_{I}\rightarrow 0$, the frequency of the harmonic
oscillator becomes $\omega_{0}=\omega_{R}$. However, the most
interesting case is that of highly excited quasinormal modes for
which $\omega_I \gg \omega_R$ and thus the frequency of the
harmonic oscillator becomes $\omega_{0}=\omega_{I}$. Furthermore,
Maggiore proposed that if one wants to solve or at least alleviate
problems that were raised by the Hod's proposal one has to employ
the $\omega_{0}$ rather than $\omega_{R}$ since in order to derive
the quantum spectrum of a black hole using its quasinormal modes,
the black hole has to be treated as a collection of damped
harmonic oscillators. In this framework, we consider the
transition $n\longrightarrow n-1$ for  a Kerr black hole. Since we
are interested in highly excited black holes, i.e. $n$ is large,
the proper frequency is now $\omega_{0}=\omega_{I}$ and thus the
absorbed energy using equations (\ref{quasi2}) and (\ref{efftemp})
is
\bea
\Delta M&=&\hbar \left[(\omega_{0})_{n} -(\omega_{0})_{n-1}\right]\label{mass1}\nn\\
&=&\hbar \left[(\omega_{I})_{n} -(\omega_{I})_{n-1}\right]\label{mass2}\\
&=&-4\pi \hbar T_{0}\label{mass3} =2\pi \hbar
T_{H}^{Sch}
\label{mass4}~.
\eea
This change in the black hole mass
will create a change in the black hole area of the form
\be
\Delta
A= 32\pi M\Delta M
\label{area1}
\ee
and substituting the change of black hole mass as given by equation (\ref{mass4}), the change
in the black hole area becomes
\be
\Delta A= 8\pi \hbar =8\pi
l_{p}^{2}
\label{quantum}
\hspace{1ex.}
\ee
A couple of comments are in order here. First, our result for the Kerr black hole
is in full agreement with that for the Schwarzschild black hole
given by Maggiore. Second, we have managed to derive a universal
area quantum, i.e. independent of the parameters that characterize
the Kerr black hole. Therefore, the concept of universality for
the area quantum has from now on a twofold meaning. On one hand,
it means that the quantum of the area spectrum is independent of
the black hole parameters and on the other hand, it means that it
is the same for the Schwarzschild and  Kerr black hole. It should
be stressed that the two meanings are interwoven since the first
statement in the limit $a\rightarrow 0$ (which reduces the Kerr
black hole to the Schwarzschild black hole) leads us directly to
the second one, and the other way around. It is noteworthy that
the change in the area of Kerr black hole (\ref{area1}) is that of
the Schwarzschild black hole. The reason for that is the fact that
we are interested in highly damped quasinormal modes where as
stated before $\omega_I \gg \omega_R$. This condition implies that
$M^{2}\gg J$ and therefore the angular part in the formula for the
horizon area change can be neglected. The same condition holds for
the effective temperature (\ref{efftemp}) of the quasinormal
frequency spectrum (\ref{quasi2}). It seems that the relaxation
time $\tau =\omega_{I}^{-1}$ is adequate for the damping to ``wash
out'' the change in the angular momentum ($\Delta J$) but not the
change in the mass ($\Delta M$).
\\

\par\noindent
Let us now try to derive the quantized area spectrum of the Kerr
black hole employing Kunstatter's method. Implementing the first
law of black hole thermodynamics (\ref{kerr1law}), the
adiabatically invariant integral (\ref{adiabatic}) is now given as
\be I= \int \frac{dM-\Omega dJ}{\omega} \label{adiabatic2}
\hspace{1ex}. \ee At this point one has to clarify what the
frequency $\omega$ in the denominator should be. For the case of a
harmonic oscillator, we claimed that this frequency is the
vibrational frequency that corresponds to the system's energy $E$
for which under a slow variation of a parameter which is related
to the energy, a small variation $dE$ in the energy was created
and the quantity $E/\omega$ is an adiabatic invariant. Following
Maggiore's proposal the perturbed black hole is treated as a set
of harmonic oscillators. In the context of this correspondence,
one has to define the cause for the small variations in the mass
($\Delta M$) and the angular momentum ($\Delta J$) of the Kerr
black hole. According to our previous syllogism, it is evident
that for the case of black holes it is the transitions of type
$n\longrightarrow n-1$, where $n\gg 1$, which  make the black hole
mass and angular momentum vary slowly and thus alter the entropy
of the black hole  through the first law of black hole
thermodynamics. Therefore, the small variations in the mass and
angular momentum of the black hole stem from the transitions and
for this reason the frequency $\omega$ should be the one that
corresponds to the absorbed energy given by equations
(\ref{mass1}) and (\ref{mass4}), i.e. the transition frequency
\bea
\omega&=&\left[(\omega_{I})_{n} -(\omega_{I})_{n-1}\right]\\
&=&2\pi  T_{H}^{Sch}\label{frequency}
\hspace{1ex}.
\eea
Therefore, the adiabatic invariant for the Kerr black hole is now
written as
\bea
I&=&\int \frac{dM -\Omega dJ}{\left[2\pi T_{H}^{Sch}\right]}\label{adiabkerr1}\\
&=&\left[2M^{2} +2\sqrt{M^4 -J^2}\right.\nn\\
&-&\left. 2M^2 \log\left({M^2 +\sqrt{M^4 -J^2}}\right)\right]
\hspace{1ex}.
\eea
Using the expression for the
Kerr black hole horizon area (\ref{area}), the adiabatic invariant
is rewritten as
\be
I= \left[\frac{A}{4\pi}-2M^2
\log\left(\frac{A}{8\pi}\right)\right]
\label{adiabkerr2}
\ee
and implementing the Bohr-Sommerfeld quantization condition
(\ref{smi}), the quantized area spectrum is
\be \mathcal{A}_n=4\pi
l_{Pl}^{2}n
\label{area2}
 \ee
where $\mathcal{A}_n$ is similar but not the same with the
quantum-corrected Kerr black hole horizon area due to logarithmic
corrections (see for instance \cite{Medved:2004yu}).
The difference stems to the fact that the logarithmic ``prefactor" $\alpha$ is not of
order unity but depends on the black hole mass, i.e. $\alpha \approx  M/l_{Pl}$.
The negative sign that accompanies the logarithmic ``prefactor" denotes that the logarithmic
correction is of microscopic nature. More importantly, it should be stressed that
since we are working with the highly damped quasinormal modes, i.e. in the large n limit,
the ``prefactor" transforms the logarithmic correction into the dominant term. This leads to
a non-equidistant area spectrum
\footnotemark[3]
\footnotetext[3]{We thank Allan Medved for indicating this subtle point.}.
\par\noindent
At this point a couple of comments are in order. First, after the present work Medved showed that if one takes the limit
$M^{2}\gg J$ (which we introduced for the derivation of the quantum of the area spectrum, i.e. equation (\ref{quantum}))
into account for the computation of the integral in equation (\ref{adiabkerr1}),
then one ends up with an evenly spaced area spectrum \cite{Medved:2008iq}.
Second, it is noteworthy that similar arguments for use of the imaginary part of the quasinormal frequencies
were presented by Kiselev \cite{Kiselev:2005ar}. In addition, Kiselev showed that the area spectrum
for the case of extremal Kerr black hole was identical with the one given here by equation (\ref{area2}),
while for the non-extremal Kerr black hole the results were significantly different
compared to the ones derived in the present analysis.
\\
We have succeeded in deriving the quantum of the area
\vspace{2ex}\footnoterule\vspace{3ex}\hspace{-2.4ex}
spectrum of Kerr black hole adopting the new physical interpretation for the
black hole quasinormal modes. This provides a strong evidence in support of the
correctness of Maggiore's proposal. The area quantum is characterized by
universality which has a twofold meaning:
%
($\alpha$) the area quantum of Kerr black hole horizon is independent of its
parameters, i.e. the mass $M$ and the angular momentum $J$, and
($\beta$) the area quantum of Kerr black hole horizon is identical
with the area quantum of the Schwarzschild black hole as derived
by Maggiore. In addition, it is worth noting that the derived area
quantum of Kerr black hole is the same with the one obtained by
Bekenstein who employed the concept of adiabatic invariants.
Finally, we proposed a new interpretation for the frequency of the adiabatic invariant
of a black hole that appears in  the context of Kunstatter's method.
This new interpretation was  introduced in order the concept of adiabatic invariant
to be incorporated in Maggiore's proposal. This identification combined with the new interpretation
of Maggiore led us to  obtain in this context the quantized area spectrum of the Kerr black
hole. However it failed to give an equidistant area spectrum since the limit $M^{2}\gg J$ was not employed.
%
%
\par\noindent
{\it Acknowledgments} I am indebted to Saurya Das and Michele
Maggiore for reading the draft of this work and for making
constructive comments. I am grateful to Allan J.M. Medved for helping me to clarify a very subtle point
in my computations. Thanks are also expressed to Emanuele Berti and Ricardo Schiappa
for fruitful correspondences and for bringing to my attention
useful references.


\end{document}